\documentclass[preprint,12pt,authoryear]{elsarticle}

\usepackage{amssymb}
\usepackage{amsmath}

\renewcommand{\d}{{\,\rm  d}}

\newcommand{\T}[1]{{#1}^{\sf  T}}

\newcommand{\abs}[1]{|#1|}

\newcommand{\sign}[1]{{\rm sgn}\left( #1 \right)}
\newcommand{\grad}[1]{{\rm grad}\left( #1 \right)}

\renewcommand{\det}[1]{{\rm det }( #1 )}

\newcommand{\la}{\langle}
\newcommand{\ra}{\rangle}

\newcommand{\fempty}[1]{{}}
\newcommand{\f}[1]{\mbox{$ #1 $}}

\newcommand{\mybold}[1]{{\boldsymbol{#1}}}

\newcommand{\fd}{\mybold{ d}}
\newcommand{\fe}{\mybold{ e}}

\newcommand{\fg}{\mybold{ g}}

\newcommand{\fn}{\mybold{ n}}

\newcommand{\fx}{\mybold{ x}}

\newcommand{\fA}{\mybold{ A}}
\newcommand{\fB}{\mybold{ B}}
\newcommand{\fC}{\mybold{ C}}

\newcommand{\fF}{\mybold{ F}}

\newcommand{\fK}{\mybold{ K}}

\newcommand{\fQ}{\mybold{ Q}}
\newcommand{\fR}{\mybold{ R}}

\newcommand{\ffA}{\mathbb{ A}}
\newcommand{\ffB}{\mathbb{ B}}
\newcommand{\ffC}{\mathbb{ C}}
\newcommand{\ffD}{\mathbb{ D}}
\newcommand{\ffE}{\mathbb{ E}}

\newcommand{\ffG}{\mathbb{ G}}

\newcommand{\ffI}{\mathbb{ I}}

\newcommand{\ffK}{\mathbb{ K}}
\newcommand{\ffL}{\mathbb{ L}}
\newcommand{\ffM}{\mathbb{ M}}

\newcommand{\ffP}{\mathbb{ P}}

\newcommand{\ffT}{\mathbb{ T}}

\newcommand{\ffV}{\mathbb{ V}}

\newcommand{\fchi}{\mbox{\boldmath $\chi$}}

\newcommand{\fsigma}{\mbox{\boldmath $\sigma$}}

\newcommand{\feps}{\mbox{\boldmath $\varepsilon $}}
\newcommand{\fgamma}{\mbox{\boldmath $\gamma $}}

\makeatletter 
\newcommand{\Spvek}[2][r]{%
  \gdef\@VORNE{1}
  \left(\hskip-\arraycolsep%
    \begin{array}{#1}\vekSp@lten{#2}\end{array}%
  \hskip-\arraycolsep\right)}

\def\vekSp@lten#1{\xvekSp@lten#1;vekL@stLine;}
\def\vekL@stLine{vekL@stLine}
\def\xvekSp@lten#1;{\def\temp{#1}%
  \ifx\temp\vekL@stLine
  \else
    \ifnum\@VORNE=1\gdef\@VORNE{0}
    \else\@arraycr\fi%
    #1%
    \expandafter\xvekSp@lten
  \fi}
\makeatother 

\newcommand*\diff{\mathop{}\!\mathrm{d}}

\newcommand{\ffzero}{{\mathpalette\makebbNull\relax}}
\newcommand{\makebbNull}[2]{%
	\raisebox{\depth}{\scalebox{0.75}[-1]{$\mathsurround=0pt#1\mathbb{O}$}}%
}

\renewcommand*\vec[1]{\underline{#1}}
\newcommand*\mat[1]{\underline{\underline{#1}}}

\newcommand*\tind[1]{{\text{#1}}}

\newcommand*\vectspace[1]{\mathcal{#1}}

\newcommand*\sSym{\vectspace{\text{Sym}}}

\newcommand*\sD[1]{\vectspace{D}^{#1}}

\newcommand*\cub{\textrm{C}}

\newcommand*\SOthree{\text{\it SO(3)}}

\newcommand\ffPsph{\ffP^{\circ}}
\newcommand\ffPdev{\ffP'}
\newcommand{\ffIS}{\ffI^{\rm S}}

\newcommand*\strain{\varepsilon}
\newcommand*\fstrain{\feps}

\newcommand*\stress{\sigma}
\newcommand*\fstress{\fsigma}

\newcommand*\eavg[1]{\la#1\ra}
\newcommand*\eff[1]{\bar{#1}}

\newcommand*\fstraineff{\eff{\fstrain}}
\newcommand*\fstresseff{\eff{\fstress}}

\newcommand*\ffCvoigt{\ffC_+}
\newcommand*\ffCreuss{\ffC_-}
\newcommand*\ffCeff{\eff{\ffC}}

\newcommand{\flambda}{\mbox{\boldmath \f{\lambda}}}
\newcommand{\fone}{\mbox{\boldmath \f{1}}}
\newcommand{\fmu}{\mbox{\boldmath \f{\mu}}}

\newcommand*\fepseff{\eff{\fstrain}}
\newcommand*\fsigmaeff{\eff{\fstress}}

\usepackage{upgreek}
\usepackage{siunitx}
\usepackage{dashbox}

\usepackage{placeins}

\newcommand\tstress{\upsigma}
\newcommand\tstrain{\upvarepsilon}

\newcommand\diag{\operatorname{diag}}

\usepackage[nameinlink, capitalize, noabbrev]{cleveref}

\journal{Mechanics of Materials}

\begin{document}

\begin{frontmatter}


\title{Diffraction Stress Factors Calculated Using a Maximum Entropy Method}

\author[1]{Maximilian Krause}
\author[2]{Nicola Simon}
\author[3]{Claudius Klein}
\author[3]{Thomas Böhlke\corref{cor1}}
\ead{thomas.boehlke@kit.edu}
\author[2]{Jens Gibmeier}

\cortext[cor1]{Corresponding author}

\affiliation[1]{
	organization={Institute of Engineering Mathematics, University of Duisburg-Essen},
	adressline={Universitaetsstr. 2},
	postcode={45141},
	city={Essen},
	country={Germany}
}

\affiliation[2]{
	organization={Institute for Applied Materials -- Materials Science and Engineering, Karlsruhe Institute of Technology (KIT)},
	adressline={Engelbert-Arnold-Str. 4},
	postcode={76131},
	city={Karlsruhe},
	country={Germany}
}

\affiliation[3]{
	organization={Institute of Engineering Mechanics, Chair for Continuum Mechanics, Karlsruhe Institute of Technology (KIT)},
	adressline={Kaiserstraße 10},
	postcode={76131},
	city={Karlsruhe},
	country={Germany}
}

\title{} 




\begin{abstract}
		Diffraction-based stress analysis of textured materials depends on understanding their elastic heterogeneity and its influence on microscopic strain distributions, which is generally done by using simplifying assumptions for crystallite interactions to calculate tensorial stress factors or in the case of very strong textures, by considering the material phase as a single crystal (crystallite group method). In this paper, we apply the micromechanical Maximum Entropy Method (MEM) to this purpose, which marks its first use for materials with texture. The special feature of this approach is a native parametrization by the effective stiffness of the material, which allows the approach to be tailored to a macroscopically measurable sample property. We perform example stress analyses of cold-rolled copper, finding through validation with full-field simulations that the MEM yields accurate local strains even for materials with extremely sharp textures. In an example stress analysis of mildly textured cold-rolled ferritic steel, the accuracy of the approach compares favorably to the established Voigt, Reuss and self-consistent Eshelby-Kröner approaches. Compared to the latter, the method is also numerically efficient to calculate.
\end{abstract}






\end{frontmatter}


\section{Introduction}
Since experiments done by \citet{lester_behavior_1925}, it is known that residual stresses influence the diffraction behaviour of metals. \cite{macherauch_sinpsi-verfahren_1961} proposed the $\sin^2(\psi)$ technique to analyze stresses via diffraction which is based on the assumption of texture-free materials without micro-eigenstrains. To analyze stresses with this technique, the material's diffraction elastic constants (DECs) must be known. If the material is textured, the material behaviour is more complex, and instead of DECs, the stress factors introduced by \citet{dolle1978einfluss} are used. These factors can be measured by performing diffraction experiments with known stress states, but can also be calculated using various theoretical models. For example, \cite{behnken_berechnung_1991} discuss stress factor calculation using the well-established Voigt, Reuss and Eshelby-Kröner (or self-consistent) models. A common feature of these models is an assumption for crystallite interactions which necessarily implies an estimate of the effective stiffness. As the estimated stiffness is not necessarily identical to the experimentally available sample stiffness, the choice of crystallite interaction introduces an additional source of error.

\citet{krause_determination_2025} used the micromechanical Maximum Entropy Method (MEM) to calculate DECs for non-textured polycrystals. The MEM, originally due to \cite{kreher_internal_1989}, is an approximation method for local field statistics which uses macroscopic properties of the material in lieu of explicit models of the microstructure. In the stress analysis context, if accurate measurements of the sample stiffness are known, the MEM resolves stiffness discrepancies between the sample and the model. 
In this paper, the MEM approach to DEC calculation is extended to stress factor calculation for polycrystals with texture.

The manuscript is structured as follows. In \cref{theoretical background}, conventions of stress factor calculation and texture representation are established.  \cref{micromechanical models} describes the derivation of the established Voigt, Reuss and self-consistent methods in this framework, followed by the textured MEM approach. As an example application, \cref{application to stress analysis of ferritic steel} contains the stress analysis of a ferritic steel sample using stress factors obtained via analytical, numerical, and experimental means. In \cref{summary and conclusions}, the results are summarized.

{\bf{Notation:}} For the formulas in this text, a symbolic tensorial notation is preferred. Vectors are written as bold lowercase letters (${\fx}$), second-order tensors as bold uppercase or greek letters (${\fA}$, ${\fsigma}$), and higher-order tensors as blackboard bold letters (${\ffC}$). The dyadic product is written ${\otimes}$, the dot product or full contraction as $\cdot$, and the Rayleigh product, which applies a rotation $\fQ$ to an arbitrary-order tensor ${\ffA}$, as ${\fQ\star\ffA}$. The linear map of a second-order tensor ${\fA}$ by a fourth-order tensor ${\ffC}$ is written using brackets (${\ffC[\fA]}$). Matrix-matrix products (${\ffA\ffB}$), transpositions ($\T{\ffA}$), inversions (${\ffA^{-1}}$) and the Moore-Penrose pseudoinverse (${\ffA^+}$)  are defined for fourth-order tensors by representing these tensors as linear maps on symmetric second order tensors, i.e., using ${6\times6}$-matrix representations. The symbol $\times$ denotes the tensorial Kronecker product which is used for second- or fourth-order tensors ${\fA}$, ${\fB}$, and ${\fC}$ as illustrated by the relation ${(\fA \times \fB) [\fC] = \fA \fC \T{\fB}}$.

\section{Theoretical Background}
\label{theoretical background}

\subsection{Texture Representation}
\label{texture representation}

In the following, we use ``texture'' to mean \emph{crystallographic texture}, i.e., a statistical description of the orientation of the crystal lattice, as opposed to \emph{morphological texture}, which would be a statistical description of the grain shape orientation. The crystal lattice orientation is characterized at any given material point by an orthogonal second-order tensor ${\fQ}$, which maps the fixed sample coordinate system $\fe_i$ to the lattice coordinate system $\fg_i$ with 
$i=\{1,2,3\}$ via
\begin{equation}
	\fg_i = \fQ \fe_i.
\end{equation}

To characterize texture, we restrict ourselves to the single-point orientation statistic, the orientation distribution function (ODF) $f$, which is a probability density distribution on $SO(3)$, and therefore has the properties
\begin{align}
	\int_{SO(3)} f(\fQ) \d Q &= 1 \label{eq:normed}, \\
	f(\fQ) & \geq 0 \label{eq:positive},
\end{align}
where $\d Q$ is the volume element in $\SOthree$. The integral in \cref{eq:normed} is volume-normalized, such that the uniform orientation distribution is equal to one. The expected value of a tensorial orientation-dependent quantity $\ffT$ is given by the orientation average
\begin{equation}
	\eavg{\ffT} = \int_{SO(3)} \ffT(\fQ) \d Q.
\end{equation}

Two different finite-dimensional representations of the texture will be used in the following. The first is based on the idea that any material sample contains a finite number of crystallites which can each be approximated by a single constant orientation value. In our case, these crystallites are all of the same size. The material is then described by a finite number of equally weighted orientations $\fQ_i$ with 
$i=\{1,\dots,N\}$, leading to a formal representation of the ODF as a sum of Dirac delta distributions
\begin{equation}
	f(\fQ) = \sum_{i=1}^N \delta(\fQ-\fQ_i).
\end{equation}
Using this representation, all expected values can be calculated as sums
\begin{equation}
	\eavg{\ffT(\fQ)} = \sum_{i=1}^N \ffT(\fQ_i).
\end{equation}
This representation of the ODF is zero almost everywhere, making it ill-suited for reasoning about local quantities.

Another representation is based on Fourier coefficients of $f$, introduced as \emph{texture coefficients} by \citet{roe_description_1965} and \citet{bunge_zur_1965}. By using tensorial representations of $SO(3)$, \citet{adams_group_1992} reformulated these coefficients into tensorial texture coefficients. For an explicit depiction of these coefficients, a basis of harmonic functions on $SO(3)$ needs to be chosen, for which we use the convention introduced by \citet{krause_tensorial_2024}, leading to the infinite series
\begin{equation}
	f(\fQ) = \sum_{n=0}^\infty \sum_{i=1}^{2n+1} \ffV_i^n \cdot (\fQ \star \ffD_i^n).
\end{equation}
The basis tensors ${\ffD^n_i}$ are \emph{deviatoric}, i.e., fully index-symmetric and trace-free in the sense that
\begin{equation}
	\ffD^n_i[\fone] = \ffzero.
\end{equation}
The materials considered in the rest of the manuscript are of cubic symmetry. Therefore, instead of an orthonormal set of basis tensors $\ffD^n_i$ which span the $n$-th order deviatoric subspace $\sD{n}$, it is sufficient to consider only a subset spanning the cubic-deviatoric subspace $\operatorname{cub}(\sD{n})$, leading to the reduced series representation
\begin{equation}
	f(\fQ) = \sum_{n=0}^\infty \sum_{i=1}^{\dim(\operatorname{cub}(\sD{n}))} \ffV_{\cub i}^n \cdot (\fQ \star \ffD_{\cub i}^n).
\end{equation}
These cubic deviatoric tensors $\ffD_{\cub i}^n$ are listed in \cref{Basis Tensors for Texture Representation} for orders up to 10.

For numerical calculations, the Fourier series is truncated up to a finite order. For sharp textures, this truncated representation  may become negative at some orientations $\fQ$, contradicting \cref{eq:positive}. However, some theoretical results derived from a truncated Fourier series are still useful. For example, as discussed by \cite{lobos_fernandez_representation_2019}, not only the Voigt and Reuss bounds, but also the Hashin-Shtrikman bounds depend only on Fourier coefficients up to fourth order, even though truncation up to fourth order may lead to negative ODF values.

\subsection{Diffraction}
\label{diffraction}

In this section, well-known properties of diffraction are recapitulated to introduce the coordinate conventions and notation used in this work. A more thorough introduction can be found, e.g., in \cite{noyan_residual_1987}. Diffraction-based measurements use the phenomenon of Bragg diffraction, by which a beam with an incident angle of $\theta$ relative to the crystal lattice plane normal direction $\fn$  is diffracted if the lattice plane distance $D$ satisfies
\begin{equation}
	m \lambda = 2 D \sin(\theta) , \quad m \in \mathbb{N}. \label{eq:bragg}
\end{equation}
The crystal lattice plane normal for a crystallite with orientation $\fQ$ is described by Miller indices $(hkl)$ via
\begin{equation}
	\fn((hkl), \fQ)  = \fQ \star \frac{h \fe_1 + k \fe_2 + l \fe_3}{\sqrt{h^2+k^2+l^2}} . \label{eq:n_micro}
\end{equation}
From the macroscopic view, the diffraction normal is prescribed by changing the relative rotations of the beam source, sample and detector. The parameterization used in this manuscript consists of an azimuth angle $\phi$ and an altitude angle $\psi$ relative to the sample surface, such that
\begin{equation}
	\fn(\varphi, \psi) = 
	\cos(\varphi) \sin(\psi) \fe_1 + 
	\sin(\varphi) \sin(\psi) \fe_2 + 
	\cos(\psi) \fe_3.
\end{equation}

When using diffraction to determine the lattice plane distance, measurements from all crystallites for which the lattice plane normal equals the diffraction normal are averaged over. In terms of their orientations, this set of diffracting crystallites can be defined by
\begin{equation}
	g = \{\fQ : \fQ ((h k l)) \fe_3 = \fn(\varphi, \psi)\}.
\end{equation}
As the lattice plane normal is invariant to any rotation $\fQ_n$ around $\fn$, an explicit parameterization is given by
\begin{equation}
	g = \{\fQ_n(\alpha) \fQ_0((h k l), \varphi, \psi) : \alpha \in [0, 2\pi)\}. \label{eq:g_explicit}
\end{equation}
$\fQ_0$ can be any rotation which maps the crystal lattice normal vector defined by Miller indices $(hkl)$ to the vector $\fn$, written schematically as
\begin{equation}
	\fQ_0 = \fQ_{\fe_3 \rightarrow \fn(\varphi, \psi)} \fQ_{(hkl) \rightarrow \fe_3}.
\end{equation}
We use Euler angles $\varphi_1$, $\Psi$, $\varphi_2$ following the convention used by \cite{roe_description_1965}, also known as ZYZ axis convention, which is defined as
\begin{align}
	\fR(\varphi_1, \phi, \varphi_2) = \fR(\varphi_2, \fe_3) \fR(\Psi, \fe_2) \fR(\varphi_1, \fe_3). \label{eq:euler} 
\end{align}
Using these angles, a possible choice for $\fQ_0$ is
\begin{align} 
	\varphi_1 &= \varphi,\\
	\phi &= \psi -\arccos(l),\\
	\varphi_2 &= \begin{cases}
		-\arcsin\left(\frac{k}{\sqrt{1-l^2}}\right)  & \abs{l} \neq 1, \\
		0  & \abs{l} = 1.
	\end{cases}
\end{align}

In diffraction, we measure the average value of an orientation-dependent quantity $\ffT$ over the set of diffracting crystallites. By definition, for a textured sample, this $g$-average reads
\begin{align}
	\eavg{\ffT}_g = \frac{\int_g f(\fQ) \ffT(\fQ) \d Q}{\int_g f(\fQ) \d Q}. \label{eq:texture_coeff_gavg}
\end{align}

Using the truncated texture coefficient series, the calculation of $g$-averages is straightforward, if involved. The fundamental idea is to use the explicit form of $g$ in \cref{eq:g_explicit} to write the integral over $g$ as an integral over the angle parameter $\alpha$, which can be calculated explicitly after introducing the texture coefficient form of the ODF. For the denominator of \cref{eq:texture_coeff_gavg}, this procedure leads to the infinite series
\begin{equation}
	\int_{g} f(\fQ) \d Q =
	\sum_{n=0}^{\infty} \sum_{i=1}^{2n+1} \frac{1}{2n+1} \left(\ffV_i^n \cdot \fQ_{\fe_3 \rightarrow \fn} \star\ffD^n_{2n+1}\right)
	\left(\ffD^n_{2n+1} \cdot \fQ_{(hkl)\rightarrow\fe_3} \star \ffD_i^n \right). \label{eq:intensity}
\end{equation}
The respective influences of the Miller indices and the diffraction angles decompose into two scalar factors. As can also be seen, all texture coefficients influence the final result.

Writing a similar explicit result for the arbitrary function $\ffT$ in \cref{eq:texture_coeff_gavg} requires a Fourier decomposition of $\ffT$. If we were averaging over the entirety of $SO(3)$, the order of relevant texture coefficients would be limited to the polynomial degree of $\ffT$ because the harmonic basis functions are orthonormal with regard to the inner product defined via integration over $SO(3)$. This does not hold for integration over the subset $g$. For the results of $g$-averages to be accurate, we therefore need to rely on convergence of the Fourier series of ${f(\fQ) \ffT(\fQ)}$, which can be assumed if the Fourier series of $f$ converges and $\ffT$ is a polynomial function of finite degree. Even if convergence is assured, numerical issues may arise when using \cref{eq:texture_coeff_gavg} if the normalization factor ${\int_g f(\fQ) \d Q}$ is close to 0.

\newcommand{\gfin}{{g_{\tind{fin}}}}
\newcommand{\dmax}{{d_{\tind{max}}}}

Calculation of $g$-averages using the texture representation described by a finite set of orientation samples requires additional assumptions, because this representation is zero almost everywhere, in particular on the set $g$. In experimental reality, the crystallites do not form a finite set of orientation samples, because parts of each crystallite are slightly misaligned due to lattice imperfections. Furthermore, due to various instrumental imprecisions such as variations in beam wavelength, beam divergences or instrumental misalignments, diffraction happens even with slight misalignments of the diffraction normal to the lattice plane normal. Both of these effects can be approximated for the finite set of orientation samples by introducing the misalignment distance
\begin{equation}
	d = \| \fn(\varphi, \psi) - \fn(hkl, \fQ) \|.
\end{equation}
As discussed by \cite{krause_determination_2025}, this distance forms a proper metric on SO(3), as it can be derived from the metric introduced by \cite{larochelle_distance_2006} under the assumption of transversal isotropy of the diffraction condition. With the misalignment distance, the diffracting subset of finite orientations is given by
\begin{equation}
	\gfin(\varphi, \psi, (hkl)) = \{\fQ_i : d(\fn(\varphi, \psi), \fn(hkl, \fQ)) \leq \dmax \} .
\end{equation}
Averages over $\gfin$ are calculated using the subsets size $N_\text{fin}$ via
\begin{equation}
	\eavg{\ffT(\fQ)}_{\gfin} = 
	\frac{1}{N_\text{fin}} \sum^{N_\text{fin}}_{i=1} \ffT(\fQ_i). 
	\label{eq:gavg_fin}
\end{equation}
As $d_{max}$ is introduced ad-hoc, it has the role of a purely numerical parameter, with no direct relation to experimental reality. While experiments also exhibit a misalignment tolerance due to the reasons discussed above, this tolerance is already part of any texture measurement via diffraction. If obtained from an experimental measurement, an additional misalignment tolerance may compound the experimental misalignment tolerance. In practice, we minimize this effect of $\dmax$ on the calculation by choosing it as small as is necessary for $g$ contain a non-zero number of crystallites for every choice of $\varphi$ and $\psi$. 

\subsection{Stress Factors}

In diffraction-based residual stress analysis, measured lattice plane distances must be linked to residual stresses, for which micromechanical relationships must be established. In this work, only single-phase polycrystals are considered. We furthermore assume that any eigenstrains in the material, such as process-induced plastic or thermal strains, are homogeneous in the volume irradiated by the beam. The penetration depth of the beam is assumed to be large enough that surface effects can be neglected. Taking these assumptions into account, the analysis is limited to residual stresses of first kind as defined by \cite{macherauch_sinpsi-verfahren_1961}. 
In this context it is sufficient to consider the infinitesimal elastic strain $\feps$, linked to the Cauchy stress $\fsigma$ by the linear-elastic Hooke's law,
\begin{equation}
	\fsigma = \ffC[\feps].
\end{equation}
We furthermore assume the Hill-Mandel condition,
\begin{equation}
	\eavg{\fsigma \cdot \feps} = \eavg{\fsigma} \cdot \eavg{\feps}.
\end{equation}
Generally, the Hill-Mandel condition requires not only an absence of stress gradients over the irradiated area, but also specific boundary conditions, e.g., periodicity or vanishing boundary stresses. As the irradiated area interacts with heterogeneous non-irradiated material at the boundary, the Hill-Mandel condition cannot be validated from experimental measurement alone and must remain an assumption. In the limit of infinitely many equal-sized grains and statistically homogeneous stress and strain fields, the Hill-Mandel condition applies based on an argument noted by, among others, \citet[section~2.5.2]{kreher_field_1985}. This suggests that, all else being equal, the error introduced by assuming the Hill-Mandel condition tends to be smaller for more finely-grained materials. 

The Hill-Mandel condition allows for the mean energy density of the irradiated volume to be described by the effective stresses and strains
\begin{align}
	\fstresseff &= \eavg{\fstress}, \\
	\fstraineff &= \eavg{\fstrain}.
\end{align}
The macro-scale behavior is given by the effective Hooke's law
\begin{align}
	\fsigmaeff = \ffCeff[\fepseff].
\end{align}

\newcommand{\gavg}[1]{\eavg{#1}_g}

Because the material is physically linear, local and effective strains are linearly related, which can be formalized via the fourth-order strain concentration tensor
\begin{align}
	\feps(\fx) = \tilde{\ffA}(\fx)[\fepseff]
\end{align}
first defined by \cite{hill_elastic_1963}. In the context of diffraction measurements, it is sufficient to consider the average strain concentration tensor over all crystallites of one orientation $\fQ$,
\begin{align}
	\ffA(\fQ) = \eavg{\tilde{\ffA}(\fx)}_{\fQ}.
\end{align}

The local elastic strain influences the lattice plane normal distance via the relationship
\begin{align}
	\frac{D-D_0}{D_0} = \feps \cdot (\fn \otimes \fn), \label{eq:lattice_plane_normal_strain}
\end{align}
where $D_0$ is the stress-free lattice plane normal distance, which we assume to be given from experimental sources. 
Using the strain concentration tensor, the relationship between lattice plane normal distance and effective stresses resolves to
\begin{align}
	\frac{\gavg{D}-D_0}{D_0} = (\fn \otimes \fn) \cdot  \gavg{\ffA} \ffCeff^{-1} [\fsigmaeff].
\end{align}
Following \cite{behnken_berechnung_1991}, we define the stress factor
\begin{align}
	\fF(\varphi, \psi, (hkl)) = \ffCeff^{-1} \gavg{\T{\ffA}}[\fn \otimes \fn], \label{eq:stress_factor}
\end{align}
which describes the linear relationship between lattice plane normal distances and effective stresses,
\begin{align}
	\frac{\gavg{D}-D_0}{D_0} = \fF(\varphi, \psi, (hkl)) \cdot \fsigmaeff. \label{eq:basic_diffraction_stress_factor}
\end{align}
Calculating $\fF$ is thus sufficient to characterize the nonlinear relationship between the diffraction normal vector angles and the lattice plane distance. To calculate $\fF$, we require $\ffA$ and $\ffCeff$, which we calculate using micromechanical models.

\section{Micromechanical Models}
\label{micromechanical models}

\subsection{Singular Approximation}
\label{singular approximation}

In this section, established theoretical models for stress factor calculation are recapitulated using the Singular Approximation introduced by \cite{fokin_solution_1972}. The Singular Approximation substitutes the random polycrystal microstructure with a statistically similar ensemble consisting of single crystallites embedded in a reference material with stiffness $\ffC_0$. As discussed in \cref{texture representation}, we assume that the polycrystal is devoid of morphological texture, such that the statistical description of the crystallite is isotropic, i.e., spherical. Under this assumption, the substitute single crystallite microstructure is equivalent to the single spherical inclusion first solved by \citet{eshelby_determination_1957}. After applying the substitute ensemble solution to the original problem, the Singular Approximation strain concentration tensor becomes
\begin{align}
	\ffA &= \ffL \eavg{\ffL}^{-1}, \\
	\ffL &= (\ffC - \ffC_0 + \ffP_0^{-1})^{-1}.
\end{align}
Following \cite{willis_variational_1981}, we write Hill's polarization tensor $\ffP_0$ for a spherical inclusion as
\begin{align}
	\ffP_0 &= \frac{1}{4 \pi} \ffIS \int_{S^2} (\fn \otimes \fn) \times \fK_0^{-1} \d n\  \ffIS, 
\end{align}
where $\d n$ is the surface element of the sphere and $\fK_0$ is the reference acoustic tensor defined as
\begin{align}	
	\fK_0 &= (\ffI \times (\fn \otimes \fn))[\ffC_0].
\end{align}

In the specific case of an isotropic reference stiffness $\ffC_0$, the polarization tensor can be solved analytically, and the result is
\begin{align}
	\ffP_0 = \frac{1}{3 K_0 + 4 G_0} \ffPsph + \frac{3 (K_0 + 2 G_0)}{5 G_0 (3 K_0 + 4 G_0)} \ffPdev.
\end{align}
We note that even when using an isotropic reference stiffness, the Singular Approximation incorporates texture-based anisotropy through the expectation value~$\eavg{\ffL}$. A more detailed discussion of the use of the Singular Approximation framework in diffraction stress analysis was given by \cite{krause_determination_2025}. For the present work, we focus on three particular choices of $\ffC_0$. By using the limit ${\ffC_0^{-1} \rightarrow \ffzero}$, we recover the Voigt model, with a strain concentration of
\begin{align}
	\ffA = \ffI^S,
\end{align}
and an effective stiffness of
\begin{align}
	\ffCvoigt = \eavg{\ffC}.
\end{align}
In the micromechanical literature, the Voigt model is generally called the \emph{Voigt bound}, as it represents an upper limit to the effective strain energy density. While this bound on the effective energy also implies a bound on the effective stiffness, we know of no similar proof for bounds on the stress factors. Therefore we refer to the Voigt \emph{model} to avoid ambiguity. The Voigt model stress factor is given by
\begin{align}
	\fF_+ = \ffCvoigt^{-1} [\fn \otimes \fn].
\end{align}
When using the Voigt model, the lattice plane normal distance is linear in $\sin^2(\psi)$, as in the non-textured case. As the stress factor depends on the texture only through $\ffCeff$, only texture coefficients up to fourth order influence the stress factor.

Assuming a vanishing reference stiffness ${\ffC_0 \rightarrow \ffzero}$ leads to the Reuss model, with the strain concentration
\begin{align}
	\ffA = \ffC^{-1} \ffCreuss
\end{align}
and an effective stiffness of
\begin{align}
	\ffCreuss = \eavg{\ffC^{-1}}^{-1}.
\end{align}
The Reuss stress factor is
\begin{align}
	\fF_- = \gavg{\ffC^{-1}}[\fn \otimes \fn].
\end{align}
As $\ffC^{-1}$ is a fourth-order polynomial in $\fQ$, an explicit calculation in terms of texture coefficients is possible, which is given in \cref{Texture Coefficient Form of Reuss Stress Factors}.

The Voigt and Reuss models result from extremal choices of the reference stiffness which are not generally realistic. A more sophisticated method is the self-consistent (SC) method, which is in the diffraction context often called the Eshelby-Kröner model. The reference stiffness $\ffC_0$ is chosen to be equal to the effective stiffness $\ffCeff$. From the general homogenization relation
\begin{align}
	\ffCeff = \eavg{\ffC \ffA}, \label{eq:homogenization}
\end{align}
a nonlinear equation for $\ffCeff$ follows. For non-textured cubic polycrystals, \cite{willis_variational_1981} showed that this equation can be simplified to a scalar cubic polynomial, which can be solved analytically. For arbitrarily textured polycrystals, $\ffC_0$ becomes anisotropic, meaning that both $\ffP_0$ and $\eavg{\ffL}$ must be calculated by numerical integration, and the overall nonlinear equation must therefore be solved numerically as well.

To solve \cref{eq:homogenization}, we use the effective stiffness form given by \cite{walpole_bounds_1966},
\begin{align}
	\ffCeff_\tind{SC} = \eavg{\ffL} - \ffCeff_\tind{SC} + \ffP_0^{-1},
\end{align}
which can be used as a fixed-point iteration reading
\begin{align}
	\ffCeff^{n+1}_\tind{SC} = \eavg{\ffL(\ffCeff_\tind{SC}^{n})} - \ffCeff_\tind{SC}^{n} + \ffP_0^{-1}(\ffCeff_\tind{SC}^{n}).
\end{align}
A suitable starting value is for example the approximation by \cite{hill1952elastic},
\begin{align}
	\ffCeff_\tind{SC}^0 = \ffCeff_\tind{Hill} = \frac{1}{2}(\ffC_+ + \ffC_-).
\end{align}

The numerical expense involved in this computation is significantly higher than for the Voigt and Reuss approaches, particularly the polarization tensor integration over $S_2$ and the texture-weighted integration over ${SO(3)}$ which need to be calculated in every iteration step. 

The numerical expense involved in this computation is significantly higher than for the Voigt and Reuss approaches, particularly the polarization tensor integration over $S2$ and the texture-weighted integration over $SO(3)$ which need to be calculated in every iteration step. As the compression modulus is homogeneous for cubic materials, the harmonic basis of \cite{krause_tensorial_2024} can be used to reduce the number of unknown components of the effective stiffness to 15, reducing the computational effort.  Once $\ffCeff$ is calculated, the stress factors can be calculated without solving a nonlinear equation by
\begin{align}
	\fF_\tind{SC} = \ffCeff_\tind{SC} \eavg{\ffL}^{-1} \gavg{\ffL}[\fn \otimes \fn],
\end{align}
which again requires numerical integration.

\subsection{The Maximum-Entropy Method}
\label{the maximum-entropy method}

The Maximum-Entropy Method (MEM) uses known information about the macroscopic properties of the sample to give an estimate for the statistics of local fields. Fundamentally, the problem of calculating local fields from effective properties is the inverse of the approach used in \cref{singular approximation}, which assumed simplified local fields to calculate effective properties. This inverse problem is underdetermined, which the MEM solves by assuming that the strain field in the sample tends towards a maximum of information-theoretic entropy. Information-theoretic entropy, introduced originally by \citet{shannon_mathematical_1948}, quantifies the amount of random information that is associated with an ensemble. The physical situation modeled by the ensemble, such as strain fields in a sample, requires a fixed amount of information to be fully described. As the amount of information that is random increases, the amount that is determined a priori decreases. Following \citet{jaynes_statistical_1963}, we therefore say that a maximum of information-theoretic entropy corresponds to a minimum of unfounded assumptions.

In applying the MEM to micromechanics, we follow the general approach of \cite{kreher_internal_1989}. The quantity of interest is the joint probability distribution of local strains and material properties, which in case of a single-phase polycrystal is described by $p(\fstrain, \fQ)$. The distribution with maximal entropy is formally found by solving the optimization problem
\begin{align}
	\max_p S = \max_p \int_{SO(3)} \int_{\sSym}  p(\fstrain, \fQ) \ln\left(\frac{p(\fstrain,\fQ)}{p_0}\right) \d \varepsilon \d Q . \label{eq:maximization_problem}
\end{align}

To ensure that the solution is compatible with applied loads and the micro- and macroscopic material properties, we apply constraints to the optimization problem. These are the definitions of effective stresses and strains,
\begin{align}
	\eavg{\fstrain} &= \fstraineff, \label{eq:strain_constraint} \\
	\eavg{\ffC[\fstrain]} &= \ffCeff[\fstraineff], \label{eq:stress_constraint}
\end{align}
and the Hill-Mandel condition,
\begin{align}
	\eavg{\fstrain \cdot \ffC[\fstrain]} = \fstraineff \cdot \ffCeff[\fstraineff]. \label{eq:energy_constraint}
\end{align}
We assume the ODF is known, leading to the additional function-valued constraint
\begin{align}
	\int_{\sSym} p(\fstrain, \fQ) \d \varepsilon = f(\fQ). \label{eq:texture_constraint}
\end{align}

The constraints are incorporated into the optimization via Lagrange multipliers. For the texture constraint in \cref{eq:texture_constraint}, as discussed by \cite{krause_maximum-entropy_2020}, this formally requires that the ODF is a square-integrable function with a corresponding Lagrange multiplier $\lambda_f$, such that the dot product reads 
\begin{multline}
	\lambda_f(\fQ) \cdot \left(\int_{\sSym} p(\fstrain, \fQ) \d \varepsilon -f(\fQ)\right) =\\
	\int_{SO(3)} \lambda_f(\fQ)  \int_{\sSym}   p(\fstrain, \fQ) - f(\fQ)  \d \varepsilon \d Q.
\end{multline}

The modified optimization problem reads
\begin{multline}
	\max_p \int_{SO(3)} \int_{\sSym}  \\  p(\fstrain, \fQ) \left( \lambda_f(\fQ) + \flambda_\strain \cdot \fstrain +  \flambda_\stress \cdot \ffC[\fstrain] + \lambda_w \fstrain \cdot \ffC[\fstrain] + 1 +   \ln\left(\frac{p(\fstrain,\fQ)}{p_0}\right)\right) \\ \d \varepsilon \d Q,
\end{multline}
where the right-hand sides of the various constraints, which do not depend on $p$, have already been omitted.

A necessary condition for the maximum of the entropy is that the partial derivative of the integrand by $p$ vanishes. This condition is also sufficient, as was proven by \cite{jaynes_statistical_1963}. As the constraint terms are linear in $p$, this condition yields an exponential function. We replace the Lagrange multipliers by parameters $\fmu_\strain$, $\fmu_\stress$, $k$ and $m(\fQ)$ as discussed by \cite{krause_maximum-entropy_2020} and rearrange the terms of $p$ to recover the multivariate normal function
\begin{equation}
	p(\fstrain, \fQ) = m(\fQ) \exp\left(-\frac{1}{2} (\fstrain -\fgamma) \cdot \ffK [\fstrain - \fgamma]\right),
\end{equation}
with mean
\begin{equation}
	\fgamma = \ffC^{-1}[\fmu_\tstrain] + \fmu_\tstress \label{eq:mem_mean}
\end{equation}
and covariance
\begin{equation}
	\ffK = k \fQ \star \ffC^{-1}. \label{eq:mem_cov}
\end{equation}

The constants $\fmu_\strain$ and $\fmu_\stress$ are determined from evaluating the constraints in \cref{eq:strain_constraint,eq:stress_constraint}, yielding
\begin{align}
	\fmu_\strain &= \ffCreuss(\ffCvoigt - \ffCreuss)^+(\ffCvoigt - \ffCeff)[\fstraineff],  \label{eq:strain_constant} \\
	\fmu_\stress &= (\ffCvoigt - \ffCreuss)^+(\ffCvoigt - \ffCeff)[\fstraineff] + \fstraineff. \label{eq:stress_constant}
\end{align} 
The Moore-Penrose pseudoinverse ${}^+$ was used instead of the usual inverse because for cubic single crystal symmetry, $\ffCvoigt-\ffCreuss$ is singular in the spherical strain mode.

The scalar covariance factor $k$ follows from \cref{eq:energy_constraint} and reads
\begin{align}
	k = \frac{1}{6} \fstraineff \cdot (\ffCvoigt - \ffCeff) (\ffCvoigt - \ffCreuss)^+ (\ffCeff - \ffCreuss)[\fstraineff].
\end{align}

Finally, the function $m(\fQ)$ is determined by \cref{eq:texture_constraint}, taking the value
\begin{align}
	m(\fQ) = f(\fQ) \sqrt{\frac{ \det{\ffC}}{(2 \pi)^6}}.
\end{align}

To use the MEM for diffraction stress analysis, we insert \cref{eq:strain_constant,eq:stress_constant} into \cref{eq:mem_mean} and factor out a strain concentration tensor, reading
\begin{align}
	\ffA_\tind{MEM} = (\ffC^{-1}\ffCreuss+\ffIS)(\ffCvoigt - \ffCreuss)^+(\ffCvoigt - \ffCeff) + \ffIS.
\end{align}
Via \cref{eq:stress_factor}, the MEM stress factor follows as
\begin{align}
	\fF_\tind{MEM}  = \ffCeff^{-1} (\ffCvoigt - \ffCeff)(\ffCvoigt - \ffCreuss)^+ \gavg{(\ffCreuss\ffC^{-1}+\ffIS)}[\fn \otimes \fn] + \ffCeff^{-1}[\fn \otimes \fn].
\end{align}
The only non-constant quantity to be averaged over $g$ is $\ffC^{-1}$, which means that it is possible to rewrite the MEM stress factor in terms of the Voigt and Reuss stress factors. This form reads
\begin{align}
	\fF_\tind{MEM} = \ffCeff^{-1} (\ffCvoigt - \ffCeff)(\ffCvoigt - \ffCreuss)^+ [\ffCreuss [\fF_-]+\ffCvoigt[\fF_+]] + \ffCeff^{-1}\ffCvoigt[\fF_+].
\end{align}
The stress factors calculated using the  MEM can be understood as a linear interpolation of the Voigt and Reuss models. Unlike other interpolations such as the direct arithmetic mean discussed by \cite{neerfeld1942spannungsberechnung}, the MEM is weighted by a term which can be directly calculated using the effective stiffness. As shown by \cite{krause_determination_2025}, for the special case of non-textured cubic materials, the MEM and a direct Voigt-Reuss interpolation are equivalent, which justifies experimental approaches to the determination of scalar interpolation weights such as that by \cite{murray2015stress}. The MEM improves on interpolations with scalar weights by providing a general solution for textured materials of arbitrary single-crystal symmetry. The MEM calculation is also numerically inexpensive, especially compared with the self-consistent model.

%

\section{Validation Example: Cold-Rolled Copper}
\label{validation example: rolled copper sheets}

\subsection{Sample Properties}
\label{taylor lin simulation}

The material under consideration is pure copper with single crystal elastic constants of
\begin{align}
	C_{1111} =& \SI{170.2}{\giga\pascal}, & C_{1122} =& \SI{114.9}{\giga\pascal}, & C_{1212} =& \SI{61.0}{\giga\pascal}, \label{eq:copper}
\end{align}
taken from \citet{simmons_single_1971}. To study extremely sharp textures, we presume that the copper has been cold-rolled into sheets, with incremental thickness reductions up to a maximum reduction of \SI{95}{\percent}. The resulting texture is obtained via a Taylor-Lin texture simulation, i.e., prescribing the macroscopic deformation process and assuming homogeneous deformation on the micro-scale \citep{taylor_plastic_1938,lin_analysis_1957}, so that the deformation of each grain can be simulated separately using a large deformation crystal plasticity material model \citep{bohlke_evolution_2001,bohlke_modeling_2003}.  We employ the \cite{kroner_allgemeine_1959} decomposition of the deformation gradient ${\grad{\fchi} = \fF_\tind{e} \fF_\tind{p}}$, splitting it multiplicatively into its elastic part $\fF_\tind{e}$ and its plastic part $\fF_\tind{p}$, which can be alternatively derived based on the assumption of isomorphic elastic ranges \cite{bertram_finite_2003}. The evolution of  $\fF_\tind{p}$ is given by a slip system based ansatz	
\begin{equation}
	\dot\fF_\text{p} \fF_\text{p}^{-1} = \sum^{N}_{\alpha=1} \dot{\gamma}_\alpha \tilde\fd_\alpha\otimes \tilde \fn_\alpha,
\end{equation}
where $\tilde\fn_\alpha$ and $\tilde\fd_\alpha$ are the slip plane normal and the slip direction of the $N$ slip systems in the undeformed reference placement. As copper has a face-centered cubic lattice, we use the twelve octahedral slip systems to describe the plastic flow. A detailed description of large deformation crystal plasticity and the octahedral slip systems is found e.g. in \cite{anand_single-crystal_2004}. For the slip rate $\dot\gamma_\alpha$, a viscoplastic approach is chosen, cf. \cite{lemaitre_mechanics_1990},	
\begin{equation}
	\dot{\gamma}_\alpha = \dot{\gamma}_0\ \sign{\tau_\alpha}\
	\Bigl< \frac{ |\tau_\alpha|-\tau_\text{C}}{\tau_\text{D}} \Bigr>^m
\end{equation}
using the projected shear stress $\tau_\alpha$ as well as the slip rate $\dot\gamma_0$, the critical shear stress $\tau_\text{C}$, the drag stress $\tau_\text{D}$ and the overstress sensitivity exponent $m$ as parameters. The values used are listed in \cref{tab:vp_model}. 	

\begin{table}[h!]
	\begin{center}
		\begin{tabular}{ |c c c c| } 
			\hline
			$\dot{\gamma}_0$ & $\tau_\text{C}$ & $\tau_\text{D}$ & $m$ \\ 
			\hline
			0.001 s$^{-1}$ & 12.0 MPa & 4.0 MPa & 20 \\ 
			\hline
		\end{tabular}	
		\caption{Viscoplastic model parameters}
		\label{tab:vp_model}
	\end{center}
\end{table}	

\begin{figure}[h!]
	\begin{minipage}{0.48\linewidth}
		\includegraphics[width=\linewidth]{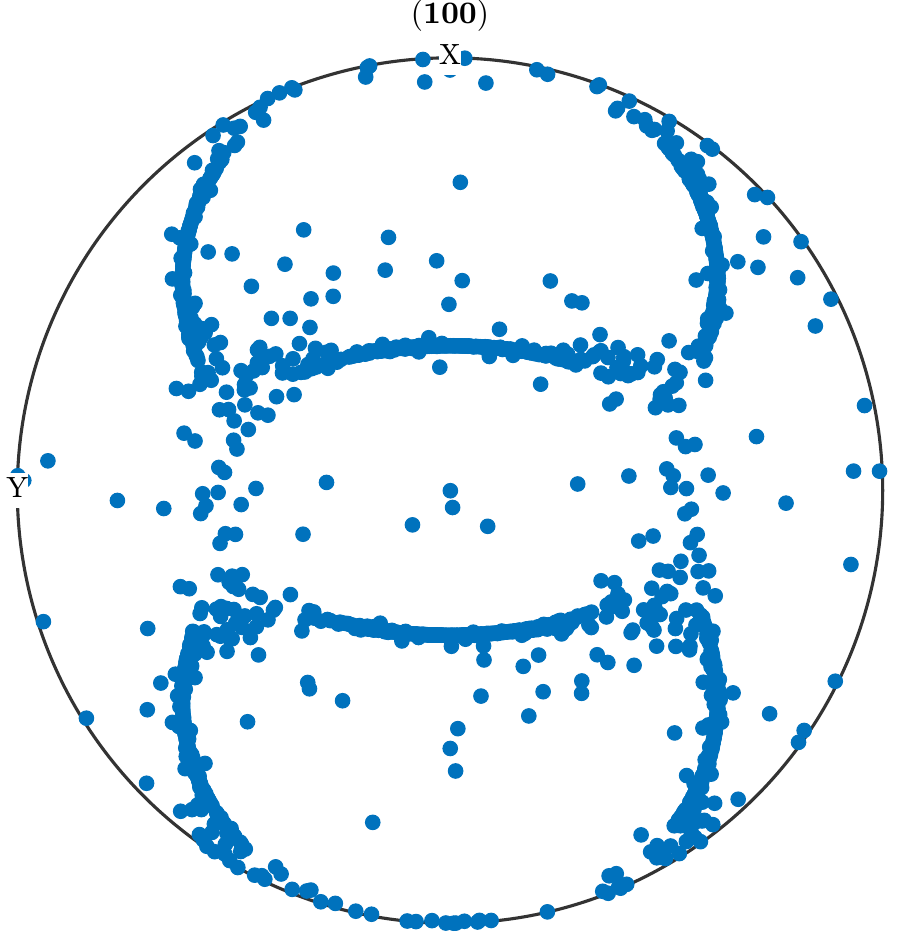}
	\end{minipage}
	\hfill
	\begin{minipage}{0.48\linewidth}
		\includegraphics[width=\linewidth]{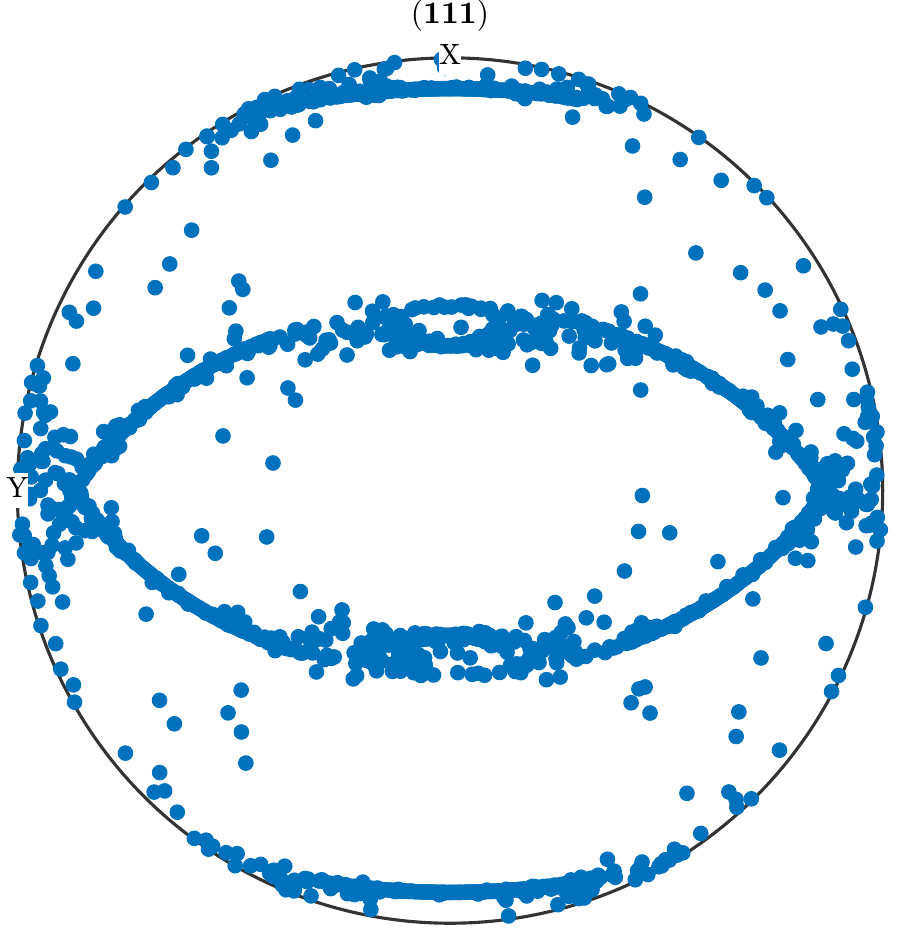}
	\end{minipage}
	\caption{(100) (left) and (111) (right) discrete pole figures for rolled copper after \SI{95}{\percent} thickness reduction, depicted with 1000 randomly chosen samples. X denotes the rolling direction, Y the direction in which no deformation occured.}
	\label{fig:polefigures}
\end{figure}

The aggregate used for the texture simulation consists of \num{10000} orientations corresponding to an initially isotropic crystallographic texture. This is achieved by sampling pseudorandomly from the Haar distribution, which is uniform on $SO(3)$, using the algorithm described by \cite{stewart_efficient_1980}.
In figure \ref{fig:polefigures}, discrete pole figures for the final state after \SI{95}{\percent} thickness reduction are shown.

\subsection{Comparing Full-Field Simulations and Analytical Approaches}
\label{Comparing Full-Field Simulations and Analytical Approaches}

\begin{figure}
	\centering
	\includegraphics[width=0.5\linewidth]{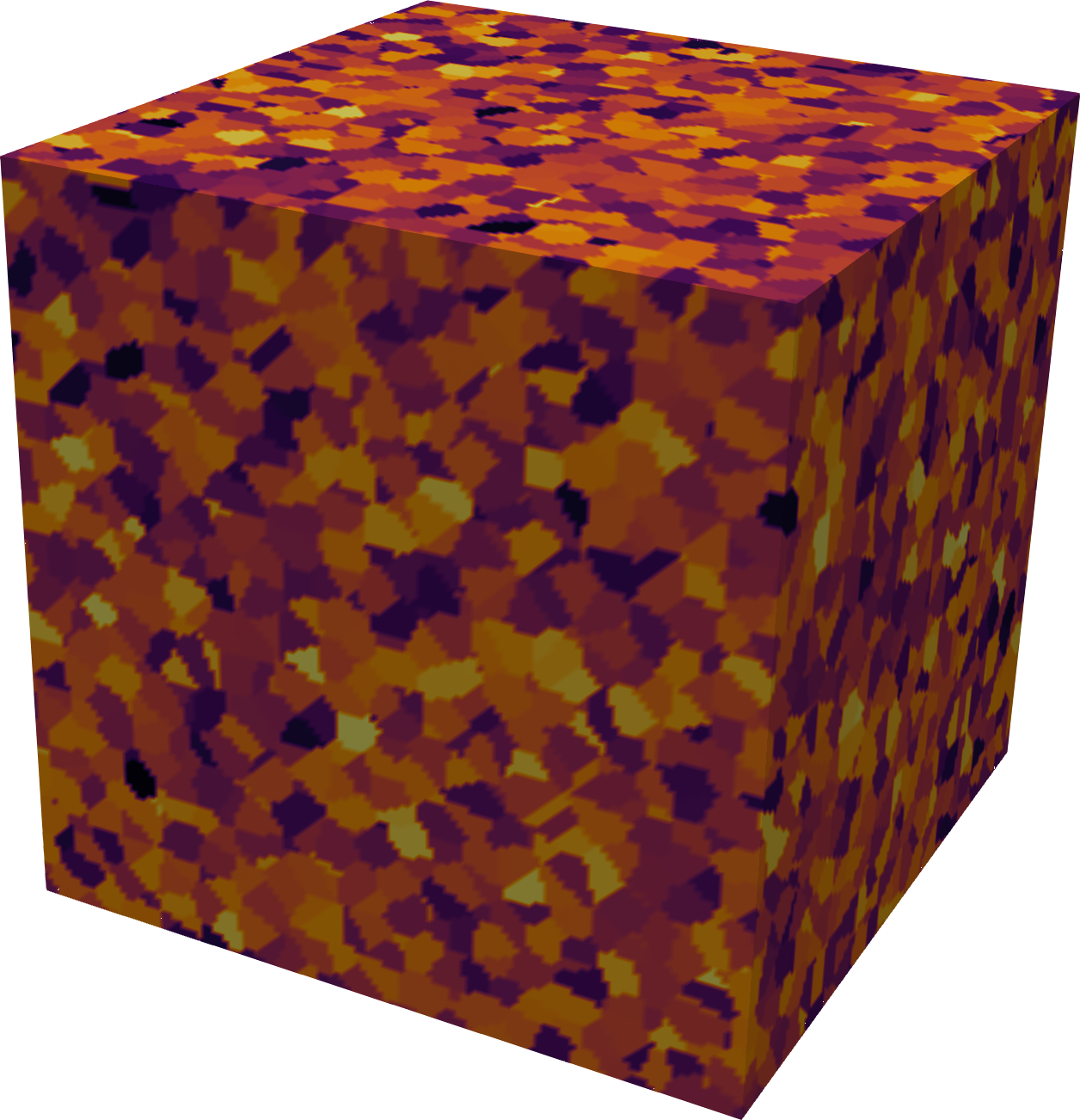}
	\caption{Microstructure used for full-field simulations, with \SI{10000} crystallites discretized using $128^3$ voxels.}
	\label{fig:microstructure}
\end{figure}

To validate the stress analysis methods in a context free of optical sources of error, we use full-field simulations of computationally generated microstructures. To this end, we simulate a volume of crystallites which are all considered to be equally involved in diffraction. As surface effects are not taken into account, this simulation is roughly equivalent to a transmission experiment using synchrotron or neutron beams. The microstructures used are generated by Laguerre tessellation as described by \cite{kuhn_fast_2020}, thus containing grains of equal size with no morphological texture. To incorporate crystallographic texture, 
we directly use the calculated orientation from the Taylor-Lin texture simulation described \cref{taylor lin simulation} by assigning the orientations randomly to the generated crystallites.
This approach differs from the algorithm by \cite{kuhn_generating_2022}, which is more accurate for elastic homogenization, but unsuitable for prescribing texture coefficients of high tensorial orders, as is required for diffraction analysis. \cref{fig:microstructure} depicts a microstructure used for the full-field simulation consisting of \SI{10000} crystallites of equal size. Based on the resolution studies performed by \cite{krause_determination_2025} for full-field simulations of non-textured polycrystals, we use a resolution of $128^3$ voxels.

To calculate local strains, we use the fast-Fourier transform approach pioneered by \cite{moulinec_numerical_1998}. We use the staggered grid discretization as described by \cite{schneider_computational_2016} to minimize Gibbs' oscillations, and solve the discretized system with a conjugated gradient approach as described by \cite{zeman_accelerating_2010}. The algorithm yields the full local strain fields resulting from periodic boundary conditions with a prescribed effective stress. From the strain field, we calculate lattice plane normal strains averaged over the set of diffracting grains $g$ using \cref{eq:gavg_fin} with $\dmax = 0.1$. We prescribe a unidirectional stress ${\fstresseff = \eff{\sigma}_i \fB_i}$ in one of six orthogonal directions $\fB_i$ in stress space. From \cref{eq:basic_diffraction_stress_factor}, the corresponding stress factor component resolves to
\begin{equation}
	F_i = \frac{\gavg{\feps \cdot (\fn \otimes \fn)}}{\eff{\sigma}_i}.
\end{equation}

\begin{figure}
	\centering
	\includegraphics[width=\linewidth]{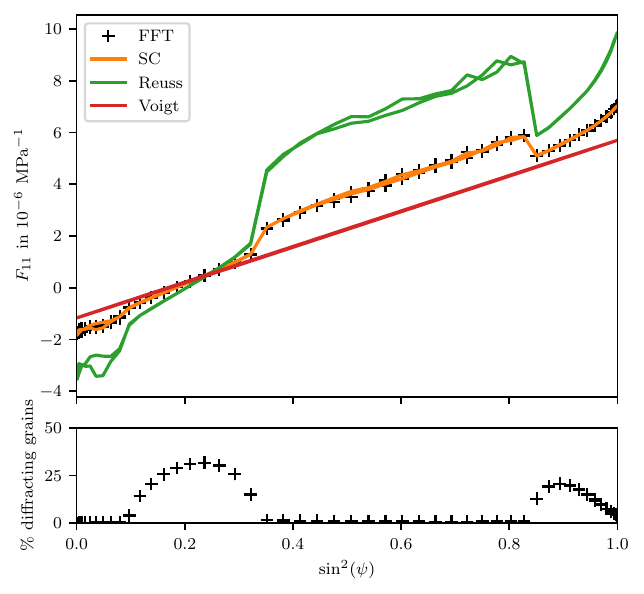}
	\caption{Stress factor component $F_{11}$ of copper rolled to \SI{95}{\percent} thickness reduction according to full-field FFT simulations, self-consistent (SC), Reuss and Voigt methods, each calculated using the same orientation samples with ${\dmax=0.1, \varphi=0^\circ}$ in the $\{311\}$ plane. Intensity shown as percentage of diffracting crystallites.}
	\label{fig:fftvsanasampled}
\end{figure}

To calculate analytical results which are directly comparable to the numerical full-field results, we follow the same procedure. Via $\ffA(\fQ)$, we directly calculate the first statistical moment of local strains in crystallites of a given orientation. We choose the same finite set of orientations as in the numerical calculation, and in averaging over $g$, we use the same misorientation tolerance $\dmax=0.1$. The resulting stress factor $F_{11}$ is shown in \cref{fig:fftvsanasampled}. A strong nonlinearity due to texture is evident.  Of the three methods shown, the self-consistent method is closest to the full-field results, which is compatible with the non-textured case discussed by \cite{krause_determination_2025}.

\begin{figure}
	\centering
	\includegraphics[width=\linewidth]{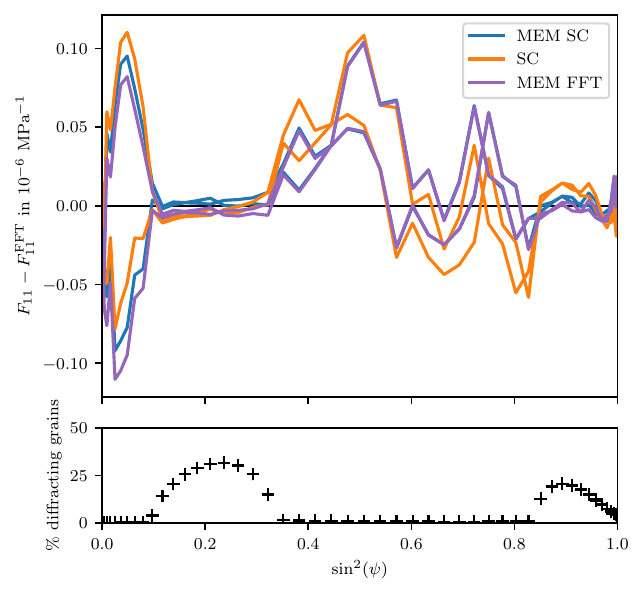}
	\caption{Stress factor component $F_{11}$ of copper rolled to \SI{95}{\percent} thickness reduction according to full-field FFT simulations, self-consistent stiffness MEM and numerical stiffness MEM, each calculated using the same orientation samples with ${\dmax=0.1, \varphi=0^\circ}$ in the $\{311\}$ plane. Only deviation from FFT results is plotted. Intensity shown as percentage of diffracting crystallites.}
	\label{fig:scmemcomparison}
\end{figure}

In \cref{fig:scmemcomparison}, a more detailed view is shown by only plotting the difference between the analytical methods and the full-field simulations. Two different MEM approaches are considered. One uses the self-consistent effective stiffness, while the other uses an effective stiffness obtained from the numerical simulation. The MEM approach using the self-consistent stiffness is not perfectly identical to the direct self-consistent calculation, as it would be in the non-textured case \citep{krause_determination_2025}, but very close. The difference between the two MEM approaches is even smaller, as the stiffnesses used are nearly identical. The intensity, shown here as the percentage of irradiated grains which is in $g$, is close to zero nearly everywhere, as expected of a sharp texture. The numerical results in areas of vanishing intensity contain a lot of noise, as too few grains are considered in the calculation. In experimental reality, it would be impossible to obtain any measurements in these areas. Considering only areas of high intensity, both MEM approaches and the self-consistent method are so close to the numerical results as to be indistinguishable and can be considered equally well-suited.

\begin{figure}
	\centering
	\includegraphics[width=\linewidth]{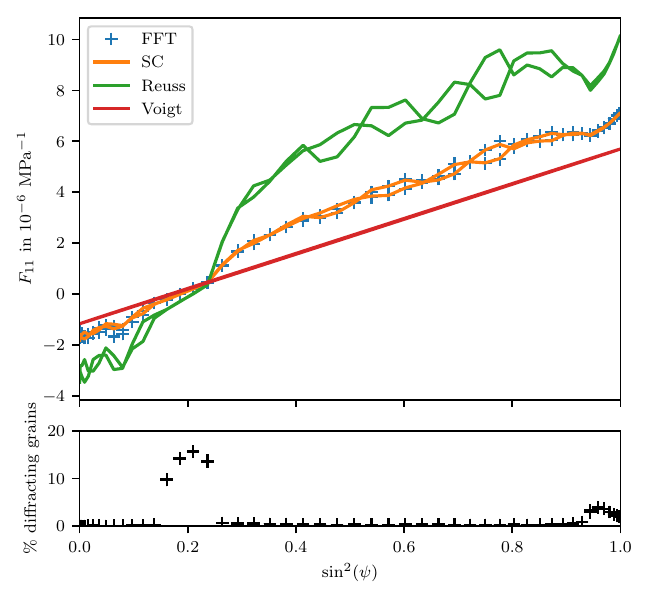}	
	\caption{Stress factor component $F_{11}$ of copper rolled to \SI{95}{\percent} thickness reduction according to full-field FFT simulations, self-consistent (SC), Reuss and Voigt methods, each calculated using the same orientation samples with ${\dmax=0.05, \varphi=0^\circ}$ in the $\{311\}$ plane. Intensity shown as percentage of diffracting crystallites.}
	\label{fig:fftvsanasampled0}
\end{figure}

In the previous diagrams, we have taken care to use the exact same texture representation for all approaches. In \cref{fig:fftvsanasampled0}, the texture representation is instead calculated by setting ${\dmax=0.05}$. As a result, both the intensity and the stress factor are changed significantly. As intensities are lower, numerical noise is more pronounced. Also, the nonlinearity has changed qualitatively to be less monotonous, again mostly due to changes in the low-intensity regions. These qualitative changes in the observed nonlinearity show that the stress factor calculations is very sensitive to errors introduced by inaccurate texture representations. We therefore conclude that while both the self-consistent model and the MEM models provide accurate stress factors in principle, problems may arise due to inaccuracies in texture representation.

\subsection{Stress Analysis Using Texture Coefficients}

In experimental practice, there is no direct access to the texture representation used by the numerical simulation. Instead, we use the tensorial texture coefficients introduced in \cref{texture representation}. This approach is equivalent to the more well-known Bunge coefficients \citep{bunge_zur_1965}. We restrict ourselves to coefficients up to order ten, which are listed in \cref{Basis Tensors for Texture Representation}. As discussed in \cref{the maximum-entropy method}, the MEM can be written as a weighted average of Voigt and Reuss approaches, which can both be written explicitly in terms of texture coefficients, unlike the self-consistent method. We therefore restrict ourselves for the following analysis to a MEM solution using a self-consistent effective stiffness tensor.

\begin{figure}
	\centering
	\includegraphics[width=\linewidth]{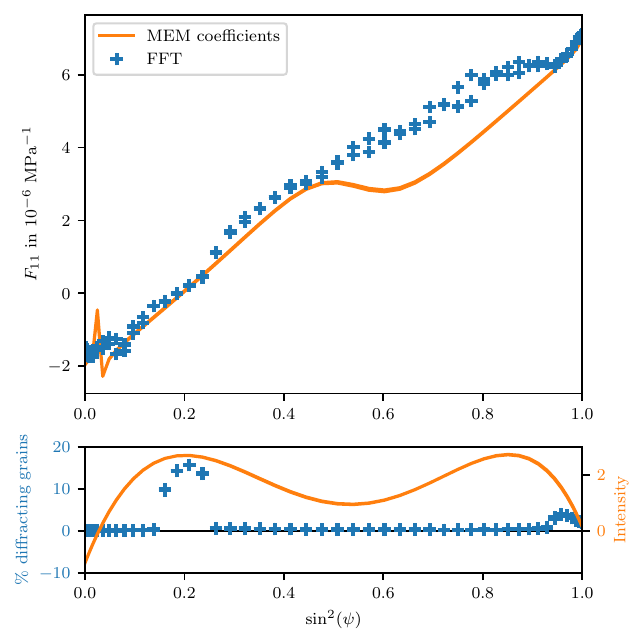}
	\caption{Stress factor component $F_{11}$ of copper rolled to \SI{95}{\percent} thickness reduction according to full-field FFT simulations with $\dmax=0.05$ and analytically calculated self-consistent stiffness MEM with tenth-order Fourier series. Calculated with ${\varphi=0^\circ}$ in the $\{311\}$ plane. 
		Intensity shown as percentage of diffracting crystallites for full-field, direct intensity value for analytical solution.}
	\label{fig:sampledmemorder100}
\end{figure}

In \cref{fig:sampledmemorder100}, the tenth-order coefficient approximation of $F_{11}$ is shown. As the intensity plot in the lower half of the image shows, tenth-order coefficients are not sufficient to represent a texture this sharp. Besides not correctly fitting the peak, the texture coefficient intensity becomes negative near ${\psi=0}$, which is not consistent with the non-negativity property of the ODF. However, in the areas where the numerical intensity is not near zero, the correspondence between analytical and numerical stress factor values is nonetheless quite good. Problems arise where the sign of the intensity changes. As the intensity is in the denominator of the stress factor, the stress factor shows a singularity at the point of zero intensity near ${\sin^2(\psi) = 0.05}$.

In texture calculation, tenth order Fourier series are a relatively rough approximation. The commonly used matlab package MTEX \citep{bachmann2010texture}, for example, uses Fourier series of order 128 in its standard settings. However, in the case of very sharp textures, increasing the Fourier order may actually increase the number of singularities, as the Gibbs' phenomenon produces ringing artifacts which cause non-negative values in functions which are zero almost everywhere. We therefore proceed with the tenth order series in attempting to perform stress analysis by applying the $\sin^2(\psi)$ method with stress factors to the full-field simulated diffraction data.



However, from an application perspective, i.e. for application to diffraction experiments, the occurrence of singularities and the associated problems are to be considered rather uncritical. In the diffraction experiments, no usable diffraction information can be obtained in these measuring directions for materials with with very pronounced crystallographic textures, so that no lattice strains can be determined for these measuring directions, hence, there is therefore no need to determine the corresponding stress factors. Based on the results in \cref{fig:sampledmemorder100}, it can be stated that from an application perspective, lattice strains can only be recorded in the range between $\sin^2(\psi) = 0.2$ and $0.95$. For these measuring directions, the analytically calculated stress factors show good agreement with those of the FFT calculation. In order to exclude the extremely high stress factors in the area of the singularities, it would be advisable, for example, to define an intensity threshold above which lattice strains actually flow into the stress evaluation.

Three diffraction measurements are simulated, with three azimuth angles $\phi \in \{0^\circ, 45^\circ, 90^\circ\}$. Each measurement point $i$ implies an equation
\begin{equation}
	(\gavg{\fstrain \cdot (\fn \otimes \fn)})_i = \fF(\varphi, \psi_i, \{hkl\}) \cdot \fstresseff \label{eq:f_system}.
\end{equation}
We concatenate these to a linear equation system
\begin{equation}
	\vec{\varepsilon}^{nn} = \mat{F} \ \vec{\eff{\stress}}, \label{eq:f_system_vec}
\end{equation}
where ${\mat{F}}$ is an ${N \times 6}$ matrix. With six variables and hundreds of simulated measurement points, this equation system is overdetermined. We solve the system using a linear least square approach, leading to 
\begin{equation}
	\vec{\eff{\stress}} = (\T{\mat{F}} \mat{F})^{-1} \T{\mat{F}} \vec{\varepsilon}^{nn}.
\end{equation}
This approach entirely disregards the inaccuracy of numerically calculated strains in regions of near-zero intensity. Therefore, we propose introducing a weighting of measurement points by the numerical (or measured) intensity. The weighted linear least-squares approach leads to
\begin{equation}
	\vec{\eff{\stress}} = (\T{\mat{F}} \mat{W} \  \mat{F})^{-1} \T{\mat{F}} \mat{W} \  \vec{\varepsilon}^{nn},
\end{equation}
with a diagonal weight matrix
\begin{equation}
	\mat{W} = \diag{n(\varphi, \vec{\psi}, d)},
\end{equation}
where ${n}$ is the number of grains within the misalignment threshold ${d=0.05}$.

However, because filtering out near-zero values of the measured intensity does not necessary filter out sign changes of the analytical approximation, the arbitrarily large values of the stress factors near singularities may still affect the least-squares approach. As a further refinement, we exclude points where the truncated Fourier series intensity is below a threshold value ${t=0.15}$, or as a formula,
\begin{equation}
	\sum_{n\in \{4,6,8,9,10\}} \frac{1}{2n+1} \left(\ffV_\cub^n \cdot \fQ_{\fn} \star\ffD^n_{2n+1}\right)
	\left(\ffD^n_{2n+1} \cdot \fQ_{hkl} \star \ffD_\cub^n \right) < t.
\end{equation}

\begin{table}
	\centering
	\begin{tabular}{|r|c c c c c c c |}
		\hline
		& {$\sigma_{11}$} & {$\sigma_{22}$} & {$\sigma_{33}$} & {$\sigma_{12}$} & {$\sigma_{13}$} & {$\sigma_{23}$} & {$e_\sigma$ (\si{\percent})}\\
		\hline
		FFT & 100 & 0.0 & 0.0 & 0.0 & 0.0 & 0.0 & 0.0\\
		least-square & 118.2 & 17.2 & 10.2 & -6.2 & 0.3 & 0.1 & 28.4\\
		weighted & 104.1 & 3.1 & 0.7 & -1.5 & 0.1 & -0.1 & 5.6\\
		no singularities & 106.5 & 4.4 & 3.1 & -2.1 & -0.0 & 0.1 & 8.9\\
		\hline
	\end{tabular}
	\caption{Stress analysis results of rolled copper using a normal least-square fit, an intensity-weighted least-square fit, and an intensity-weighted least-square fit with singularities removed. All stresses in \si{\mega\pascal}.}
	\label{tab:stress analysis}
\end{table}

The applied stress in the full-field simulation is a unidirectional stress in $11$-direction of \SI{100}{\mega\pascal}. From the fitting procedure, we retrieve the values shown in \cref{tab:stress analysis}. The unweighted least-square approach has errors of nearly \SI{30}{\percent}. Introducing the weights reduces this error to a single-digit percentage. Removing singularities via a threshold in this case increases the error further, which suggests that errors introduced by singularities actually happen to counteract other errors in this case.

\begin{figure}
	\centering
	\includegraphics[width=\linewidth]{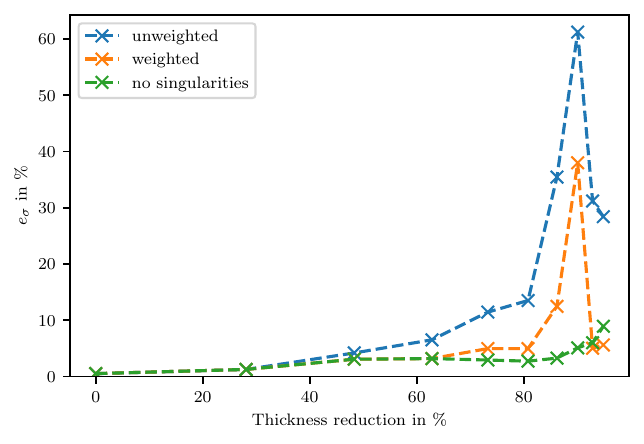}
	\caption{Relative stress analysis error compared to prescribed numerical effective stress calculated using self-consistent sitffness MEM stress factors with unweighted least-squares, weighted least-squares and singularity-filtered weighted least squares for increasingly sharp textures.}
	\label{fig:erroroversharpness}
\end{figure}

In \cref{fig:erroroversharpness}, the same procedure was repeated for other texture states throughout the rolling process. For cold-rolling up to $\SI{30}{\percent}$ thickness reduction, no non-negative intensities occur, and no weighting is necessary in the least-square analysis. At roughly $\SI{80}{\percent}$, singularities produce sizable errors in the methods in which singularities were not removed via a threshold. As the thickness reduction increases to \SI{95}{\percent}, the error of the singularity-corrected method increases significantly.

As shown in \cref{fig:erroroversharpness} and discussed in more detail by \cite{krause_determination_2025}, for non-textured copper, the error values are below \SI{1}{\percent}. Taking into account the close correspondence of full-field simulations and the MEM observed when using the same texture representation, it appears that an imperfect choice of texture representation can introduce significant errors, which can be somewhat mitigated by adjusting the least-square fitting procedure. We note that cold-rolling to a thickness of \SI{95}{\percent} represents an extreme case. In the following, we discuss an example with a significantly milder texture providing experimental data from diffraction analyses.

\section{Validation Example: Cold-Rolled Ferritic Steel}
\label{application to stress analysis of ferritic steel}

\subsection{Sample Properties}
\label{material properties}

The material under consideration is the corrosion resistant ferritic chromium steel X6Cr17 (mat.no. 1.4016). We assume the single crystal stiffness is close to that for pure iron, which is given by
\begin{equation}
	C_{1111} = \SI{230}{\giga\pascal}, \quad C_{1122} = \SI{135}{\giga\pascal}, \quad C_{1212} = \SI{117}{\giga\pascal},
\end{equation}
according to Landolt-Börnstein \citep{every_second_1992}.

\begin{figure}
	\centering
	\includegraphics[width=\linewidth]{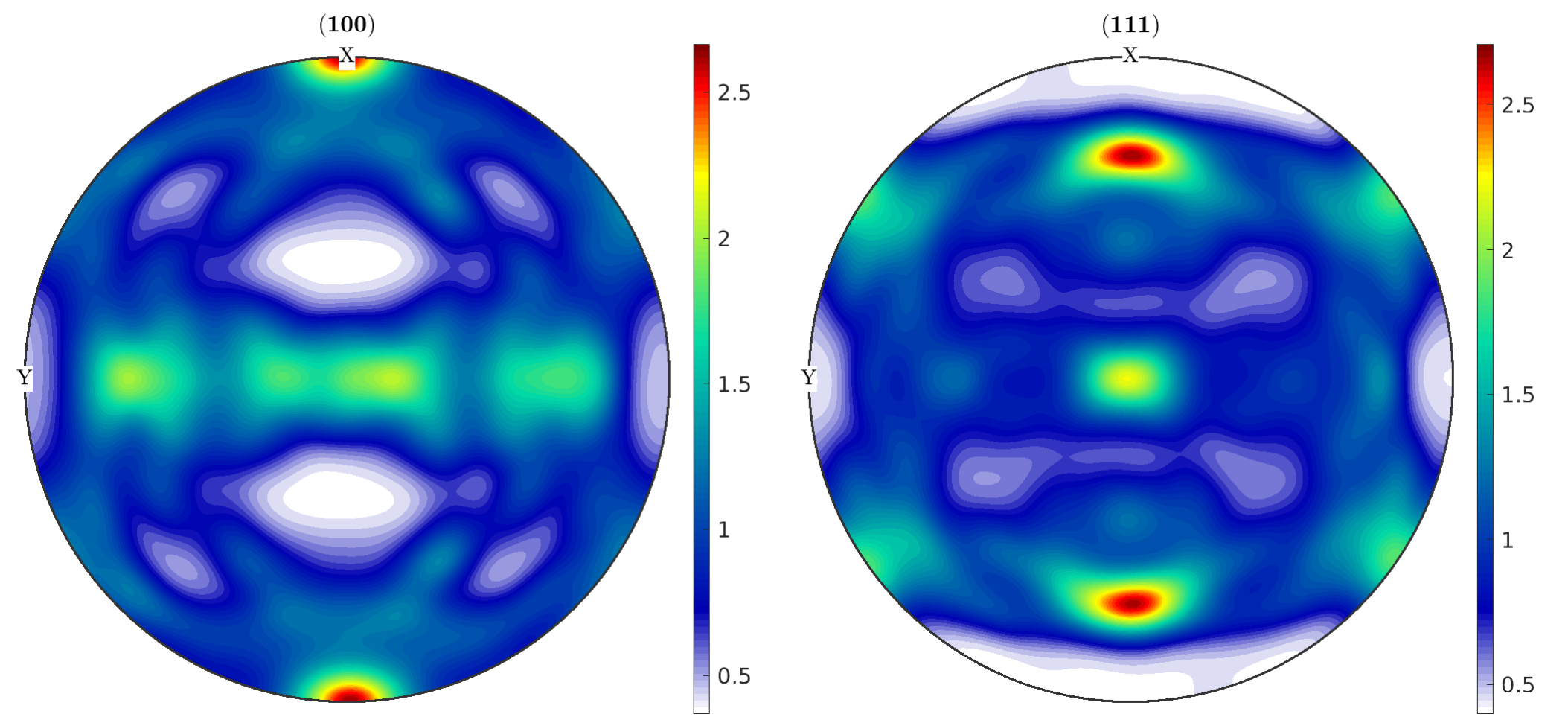}
	\caption{(100) (left) and (111) (right) pole figures recalculated on base of the experimentally determined ODF. X denotes the rolling direction, Y the transverse direction.}
	\label{fig:measured}
\end{figure}

All measurements are performed on samples cut from sheets which were cold-rolled to a final thickness of \SI{1.5}{\milli\meter}. The resulting texture was obtained based on X-ray diffraction experiments in relection geometry using a four-circle X-ray diffractometer of type Seifert PTS with \SI{1}{\milli\meter} collimated Fe-filtered Co–K$\alpha$-radiation (photon energy ${E^{\tind{CoK}\alpha} = \SI{6.93}{\kilo\electronvolt}}$, wavelength ${\lambda_{\tind{CoK}\alpha} = \SI{1.789}{\angstrom}}$. On the secondary side, a \SI{4}{\milli\meter} slit aperture was installed in front of the point detector. From the measured incomplete pole figures of the lattice planes $\{200\}\alpha$, $\{211\}\alpha$, $\{220\}\alpha$, the ODF $f(g)$ was calculated using the Matlab toolbox MTEX \citep{bachmann2010texture}. Example reconstructed pole figures are depicted in \cref{fig:measured}, showing a rolling texture which is not particularly sharp.

	\subsection{Experimental Diffraction Measurements}
	
	For experimental validation, in-situ X-ray diffraction (HEXRD) experiments were carried out in transmission geometry at beamline P21.2@PETRA III at DESY in Hamburg, Germany. We used monochromatic synchrotron X-rays with a photon energy of ${E^{HE} = \SI{100}{\kilo\electronvolt}}$ (wavelength ${\gamma^{HE} = \SI{0.12398}{\angstrom}}$). The beam size was \SI{1}{\mm} in width, \SI{1}{\mm} in height. Diffraction patterns were detected by a Varex 2D-detector of type XRD 4343CT with a $2880 \times 2880$ pixel array with a pitch size of $\SI{150}{\micro\meter} \times \SI{150}{\micro\meter}$, which was placed at a distance to the sample of \SI{1697}{\milli\meter}. The experimental setup was calibrated using a Fe-powder
	reference sample. A more detailed description of a comparable experimental setup can be found in \cite{simon2023oscillating}.
	
	The material was subjected to an uniaxial deformation, during which  diffraction patterns were continuously recorded with an acquisition time of ${\Delta t = \SI{2}{\second}}$ (sampling rate \SI{0.5}{\hertz}). Diffraction rings of the lattice planes $\{110\}$, $\{200\}$, $\{211\}$, $\{220\}$ were detected.  The rings were sectioned into 144 sections, each of which was integrated to correspond to one value of the angle $\psi$ as discussed by \cite{simon2023oscillating}.
	
	\subsection{Comparison between Full-Field, Analytical and Experimental Results}
	
	During the uniaxial deformation, the evolution of plastic strains caused heterogeneous micro-eigenstrains. As the analysis method assumes homogeneous eigenstrains in the sample, the evolution of heterogeneous micro-eigenstrains cannot be taken into account. Thus, we assume that during unloading, no further plastic deformation occurs, and use the difference in measured lattice plane distance between the final step under load and the unloaded step as the basis of our analysis. 
	
	For the material under consideration, the uniform lattice spacing for the stress-free state $d_{\{100\}}$ is a priori unknown. We use the average of the uniform lattice spacings of all analysed interference lines measured for the materials initial state, as discussed by \cite{noyan_residual_1987}. This value resolves to \SI{2.8723}{\angstrom}. The stress before unloading measures $\sigma_{11} = \SI{485}{\mega\pascal}$. Using these values, $F_{11}$ can be calculated for each measurement point via
	\begin{equation}
		F_{11} = \frac{D - D_0} {D_0 \eff{\stress}_{11}}.
	\end{equation}

	\begin{figure}[h]
		\centering
		\includegraphics[width=\linewidth]{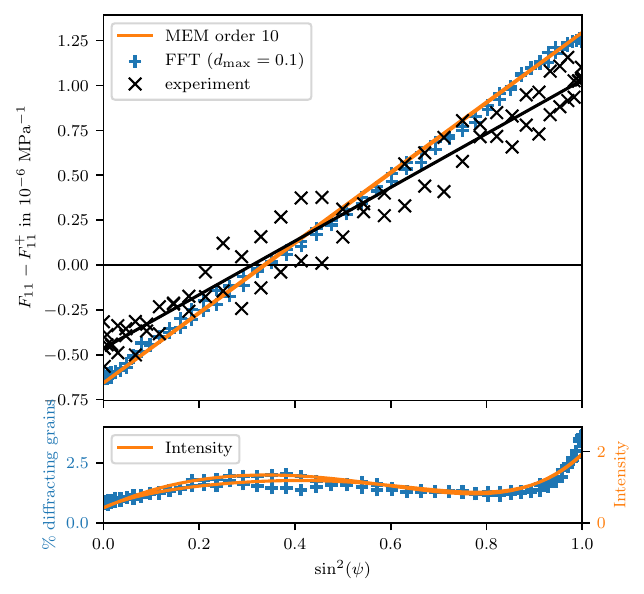}
		\caption{Stress factor component $F_{11}$ of ferritic steel X6Cr17 according to full-field FFT simulations ($\dmax=0.05$), MEM using texture coefficients up to tenth order, and experimental values with linear fit. Values given for ${\varphi=0^\circ}$ for the $\{200\}$  planes. Intensities for the FFT calculation are shown as a percentage of irradiated crystallites, while intensities for the MEM are directly based on texture coefficients.}
		\label{fig:f11minusvoigtpaper}
	\end{figure}
	
	In \cref{fig:f11minusvoigtpaper}, analytical, experimental and numerical values of $F_{11}$ for the $\{200\}$ lattice plane are compared. For each solution, the difference to an analytical Voigt solution is plotted to showcase the effect of texture. For the analytical approach, we choose a tenth-order texture coefficient approximation for the ODF, which leads to a good match of the intensity, unlike in the strongly textured case. The MEM solution using the self-consistent effective stiffness corresponds well with the numerical calculations. The experimental results agree in principle, but show pronounced scattered data. Regardless of texture, all results are expected to be linear for the $\{200\}$ lattice plane as discussed by \cite{evenschor1975rontgenographische}. Therefore, we have also plotted a least-squares linear approximation of the experimental results. This suggests that the results lie between the self-consistent and the Voigt method, of which the measurements are much closer to the self-consistent approach. 
	
	\begin{figure}[h]
		\begin{minipage}{0.5\textwidth}
			\centering
			\includegraphics[width=\linewidth]{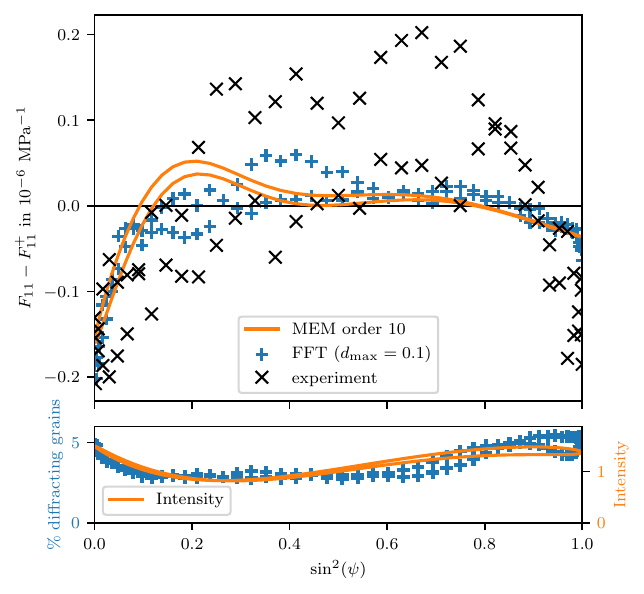}
		\end{minipage}
		\begin{minipage}{0.5\textwidth}
			\centering
			\includegraphics[width=\linewidth]{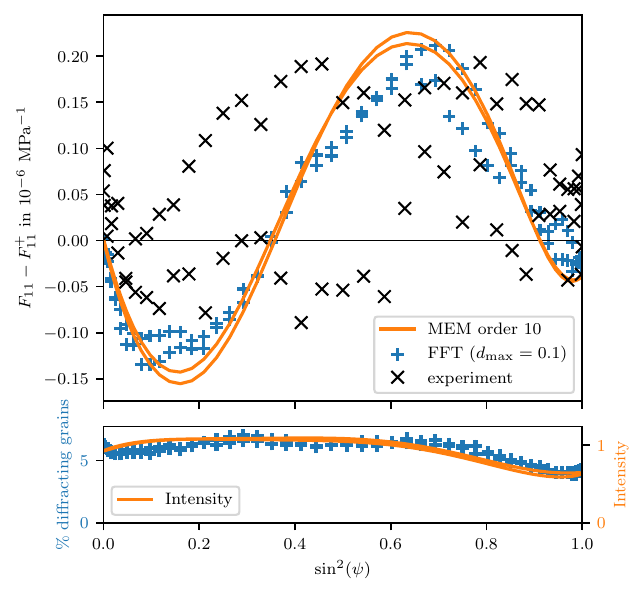}
		\end{minipage}
		\caption{Stress factor component $F_{11}$ of ferritic steel X6Cr17 according to full-field FFT simulations ($\dmax=0.05$), MEM using texture coefficients up to tenth order, and experimental values. Values given for ${\varphi=0^\circ}$ for the $\{220\}$ planes (left) and the $\{211\}$ planes (right). Intensities for the FFT calculation are shown as a percentage of irradiated crystallites, while intensities for the MEM are directly based on texture coefficients.}
		\label{fig:f11minusvoigtpaper220}
	\end{figure}
	
	Measurements for the $\{220\}$ and $\{211\}$ lattice planes are reproduced in \cref{fig:f11minusvoigtpaper220}. For these planes, the analytical methods lie more closely together. As a result, the scatter of the experimentally determined data is even more pronounced relative to the analytical differences. Therefore, no decisive statement can be made on the suitability of the various analytical methods. However, we note again a strong correspondence between Full-Field simulations and the MEM approach. The measurement for the $\{110\}$ lattice planes is qualitatively similar to that for the $\{220\}$ lattice planes, but contains even more scatter, and is therefore not depicted.

	%
	%
	
	%
	%
	%
	%
	%
	
	%
	
	\FloatBarrier
	\section{Summary and Conclusions}
	\label{summary and conclusions}
	
	In this work, a novel method of calculating stress factors for diffraction stress analysis of textured polycrystals was developed. This approach uses the Maximum-Entropy Method (MEM) to estimate local strain fields. The advantages are as follows:
	\begin{itemize}
		\item Compared to existing methods using specific assumptions for crystallite interactions, such as the Voigt, Reuss or self-consistent (Eshelby-Kröner) methods, the MEM takes the effective stiffness as an input parameter, allowing for more accurate stress analysis by incorporating measurable macroscopic properties of the sample.
		\item As the MEM can be written as a weighted interpolation of the Voigt and Reuss approaches, implementations can be based on existing code. The numerical expense of calculation is very small, especially compared to the self-consistent method.
		\item Compared to existing Voigt-Reuss-interpolation approaches for non-textured materials, the MEM can be naturally extended to textured polycrystals and uses an interpolation parameter that can be measured without diffraction, the effective stiffness.
	\end{itemize}
	
	In validating the MEM, we used a full-field simulation approach using the local strain fields of computationally generated microstructures were calculated. Besides allowing for a very precise comparison context free of instrumental effects and measurement errors, full-field simulations can  be used to directly calculate stress factors, which yielded the following insights:
	
	\begin{itemize}
		\item In our simulations for both sharply and mildly textured cubic polycrystals, the simulated stress factors were numerically indistinguishable from those given by the MEM or self-consistent method, extending the trend noted by \cite{krause_determination_2025} for texture-free polycrystals. 
		\item The MEM yielded similar results whether the self-consistent stiffness or numerically simulated effective stiffnesses were used as input, as both stiffnesses were very close, showing again that the effective stiffness characterizes the diffraction behavior.
		\item Excellent correspondence of any method to the result was only possible when using the same finite texture representation for the simulation and the analytical models. Stress analysis appears to be more sensitive to slight errors in texture representation than to slight errors in the micromechanical modeling.
	\end{itemize}   
	
	For a very sharply textured cold-rolled copper sheet, numerical full-field simulations were performed to validate the MEM. By performing stress analysis using the calculated stress factors on the numerically simulated diffraction measurements, we found that as the texture grows sharper with subsequent rolling steps, the analysis is made substantially more accurate by incorporating measurements of the intensity as weights for the fitting procedure. By a weighted least-square approach with filtering of analytical singularities, an analysis error of below \SI{10}{\percent} was possible for copper rolled to \SI{95}{\percent} thickness reduction. In this case, a coefficient approximation of the ODF of only tenth degree was used, limiting the number of nonphysical singularities.
	
	We performed experimental diffraction measurements for mildly textured ferritic steel to further validate the MEM. For stress factors calculated from the measurements, the degree of measurement noise in the $\{220\}$, $\{211\}$ and $\{110\}$ lattice plane largely overshadowed the small differences between various analytical models. In the $\{200\}$ lattice plane, a comparison with the analytical models found that a MEM model with the self-consistent effective stiffness again offers the best model out of those considered. However, the measurements suggest that a MEM model with a slightly higher effective stiffness might perform better. In light of the high levels of statistical noise, we did not attempt to fit the effective stiffness. We note that if a systematically higher or lower stiffness than the self-consistent model is observed in subsequent diffraction experiments or corroborated by sufficiently accurate measurements of the stiffness, the MEM is capable of directly incorporating these results to yield more accurate stress factors for stress analysis.
	
	\vspace{.5cm}
	\emph{Acknowledgments} The research documented in this paper was funded (BO 1466/14-1 and GI376/28-1) by the German Research Foundation (DFG) as part of project 512640977, 'Evaluation of non-linear \f{\sin^2{\psi}} - distributions in residual stress analysis based on a scale-bridging mechanical modeling'. The support by the German Research Foundation (DFG) is gratefully acknowledged. 
	
	\emph{Author Contributions}
	Conceptualization, M.K.; Methodology, M.K.; Software, M.K. and C.K.; Validation, M.K.; Formal Analysis, M.K.; Investigation, M.K., N.S. and C.K.; Resources, J.G. and T.B.; Data Curation, M.K.; Writing--Original Draft, M.K., N.S. and C.K.; Writing--Review \& Editing, M.K., N.S., C.K., J.G. and T.B.; Visualization, M.K.; Supervision, T.B.; Project Administration, T.B.; Funding Acquisition, J.G. and T.B.
	
	\emph{Conflicts of Interest}
	The authors declare no conflict of interest.

%
%
%
%

\appendix
\section{Basis Tensors for Texture Representation}
\label{Basis Tensors for Texture Representation}

As in \cite{krause_tensorial_2024}, we define an orthonormal basis of the $n$-th order deviatoric subspace $\sD{n}$ by calculating eigentensors of the rotation around the axis $\fe_3$, which is in this work the normal vector of the sample surface. For the $n$-th order space, there are $2n+1$ complex-valued eigentensors $\ffE^n_i$, grouped into pairs of complex conjugates, with the remaining real eigentensor being $\operatorname{dev}(\fe_3^{\otimes 3})$. Each complex conjugate pair can be combined into a pair of real tensors, leading to the basis tensors $\ffD^n_i$. A more detailed description of the required calculations is given in \cite{krause_determination_2025}.

With the cubic symmetry group ${C = \{\fQ_j\}}$ consisting of 24 rotations, the deviatoric basis tensors are projected onto the cubic deviatoric subspace via
\begin{equation}
	\tilde{\ffD}^n_{\cub i} = \frac{1}{24}\sum^{24}_{j=1} \fQ_j\star \ffD^n_i, \qquad \fQ_i \in C.
\end{equation}
After re-orthogonalizing $\tilde{\ffD}^n_{i}$ using the Gram-Schmidt procedure, the cubic basis tensors $\ffD^n_{\cub i}$ are retrieved. Up to order 10, all cubic deviatoric subspaces have a dimension of at most 1. Their basis tensors are
\begin{align}
	\ffD^{4}_{\cub 1} &= \frac{\sqrt{15}}{6}\ffD^{4}_{1}
	+ \frac{\sqrt{21}}{6}\ffD^{4}_{9}, \\
	\ffD^{6}_{\cub 1} &= - \frac{\sqrt{14}}{4}\ffD^{6}_{5}
	+\frac{\sqrt{2}}{4}\ffD^{6}_{13}, \\
	\ffD^{8}_{\cub 1} &= \frac{\sqrt{195}}{24}\ffD^{8}_{1}
	+\frac{\sqrt{21}}{12}\ffD^{8}_{9}
	+\frac{\sqrt{33}}{8}\ffD^{8}_{17}, \\
	\ffD^{9}_{\cub 1} &= - \frac{\sqrt{42}}{12}\ffD^{9}_{4}
	+ \frac{\sqrt{102}}{12}\ffD^{9}_{12}, \\
	\ffD^{10}_{\cub 1} &= - \frac{\sqrt{1122}}{48}\ffD^{10}_{5}
	- \frac{\sqrt{22}}{8}\ffD^{10}_{13}
	+ \frac{\sqrt{390}}{48}\ffD^{10}_{21}.
\end{align}

\section{Texture Coefficient Form of Reuss Stress Factors}
\label{Texture Coefficient Form of Reuss Stress Factors}

By using the projector decomposition of $\ffC_\cub$, we find that 
\begin{equation}
	\ffC^{-1}(\fQ) = \fQ \star \ffC_\cub^{-1} = \frac{1}{\lambda_1} \ffPsph + \frac{1}{\lambda_2} (\fQ \star \ffD_\cub^4 - \ffPsph) + \frac{1}{\lambda_3} (\ffIS - \fQ \star \ffD_\cub^4)
\end{equation}
with stiffness eigenvalues ${\lambda_1, \lambda_2, \lambda_3}$. The $g$-average yields
\begin{equation}
	\gavg{\ffC^{-1}} = \left(\frac{1}{\lambda_1} - \frac{1}{\lambda_2}\right) \ffPsph + \frac{1}{\lambda_3} \ffIS + \left(\frac{1}{\lambda_2} - \frac{1}{\lambda_3} \right) \frac{\int_g f(\fQ) \fQ \star \ffD_\cub^4 \d Q}{\int_g f(\fQ) \d Q}.
\end{equation}
The denominator of the final term is given by \cref{eq:intensity}. A similar form for the numerator is
\begin{equation}
	\int_{g} \fQ \star \ffD_\cub^{4} \f(\fQ) \diff V(\fQ) =  \fQ_{\fe_3 \rightarrow \fn} \star
	\sum_{n=0}^{\infty} \sum_{j=1}^{2n+1} 
	\frac{1}{2n+1} \ffM^{4n}_j
	[\fQ_{\fe_3 \rightarrow \fn}^{-1} \star \ffV_j^n], \label{eq:D_g_avg}
\end{equation}
with
\begin{align}
	\ffM^{4n}_j = \sum_{o=|m-n|}^{m+n}
	\left((\ffD_\cub^4 \otimes \ffD_j^n) \cdot \ffG^{4no} [\fQ_{(hkl)\rightarrow\fe_3} \star \ffD_{2o+1}^o]\right)
	\ffG^{4no}[\ffD_{2o+1}^o],
\end{align}
where $\ffG^{mno}$ is the Clebsch-Gordan tensor as defined by \cite{krause_tensorial_2024}. Notably, this innermost sum depends only on the Miller indices and can be pre-computed, which speeds up the evaluation. Generally, the Reuss stress factors depend on texture coefficients of all orders. However, if the Fourier series of the ODF converges, so does this series, and a sufficiently high-order truncated representation can in theory be arbitrarily accurate. 

\bibliographystyle{elsarticle-harv}
\bibliography{lit.bib}

\end{document}